 \documentclass[final,5p,times,twocolumn]{elsarticle}

\usepackage{amssymb}

\journal{Physica E}

\begin{document}

\begin{frontmatter}



\title{Time Resolved Control of Electron Tunnelling Times and Single-shot Spin Readout in a Quantum Dot}


\author{L. Gaudreau$^{a, b}$, A.Kam$^{a}$, J. B. Kycia$^{c}$, S. A. Studenikin$^{a}$, G. Granger$^{a}$, J. D. Mason$^{c}$, P. Zawadzki$^{a}$ and  A.S. Sachrajda$^{a}$}

\address{$^{a}$Institute for Microstructural Sciences, National Research Council, Ottawa, ON, Canada, K1A 0R6}
\address{$^{b}$Physics Department, University of Sherbrooke, QC, Canada, J1K 2R1}
\address{$^{c}$ Department of Physics and Astronomy, University of Waterloo, Waterloo, ON, Canada N2L 3G1}

\begin{abstract}

We are pursuing a capability to perform time resolved manipulations of single spins in quantum dot circuits involving more than two quantum dots. In this paper, we demonstrate full counting statistics as well as averaging techniques used to calibrate the tunnel barriers. We make use of this to implement the Delft protocol for single shot single spin readout in a device designed to form a triple quantum dot potential. We are able to tune the tunnelling times over around three orders of magnitude. We obtain a spin relaxation time of 300 $\mu$s at 10 T.

\end{abstract}

\begin{keyword}
Quantum dot; Charge detection; Full counting statistics; Spin relaxation time

Pacs: 72.25.Hg; 72.25.Rb; 73.23.Hk; 73.63.Kv

\end{keyword}

\end{frontmatter}


\section{Introduction}
\label{}

Since the first demonstration of the few electron lateral quantum dot\cite{Ciorga2000} and its subsequent combination with charge detection in a double quantum dot\cite{Elzerman2003}, there has been a lot of effort devoted to demonstrating individual DiVincenzo criteria for single spin qubits\cite{Loss1998}. In order to progress further, one will need to combine these achievements within appropriate architectures and incorporate more complex functionalities such as spin busing and error correction. In this direction, we are developing quantum dot circuits involving more than two quantum dots. Recently, we successfully demonstrated a fully tuneable triple quantum dot device\cite{Gaudreau2009}. In this paper, we use an alternative triple quantum dot design which worked only marginally as a triple quantum dot to demonstrate how we calibrate and set the tunnel barriers. Using these values we run the Delft protocol for single spin readout first demonstrated by Elzerman \textit{et al.}\cite{Elzerman2004} and extract the spin relaxation time T$_1$\cite{Amasha2008}.

\begin{figure}[htb]
\begin{center}
\includegraphics*[scale=0.7]{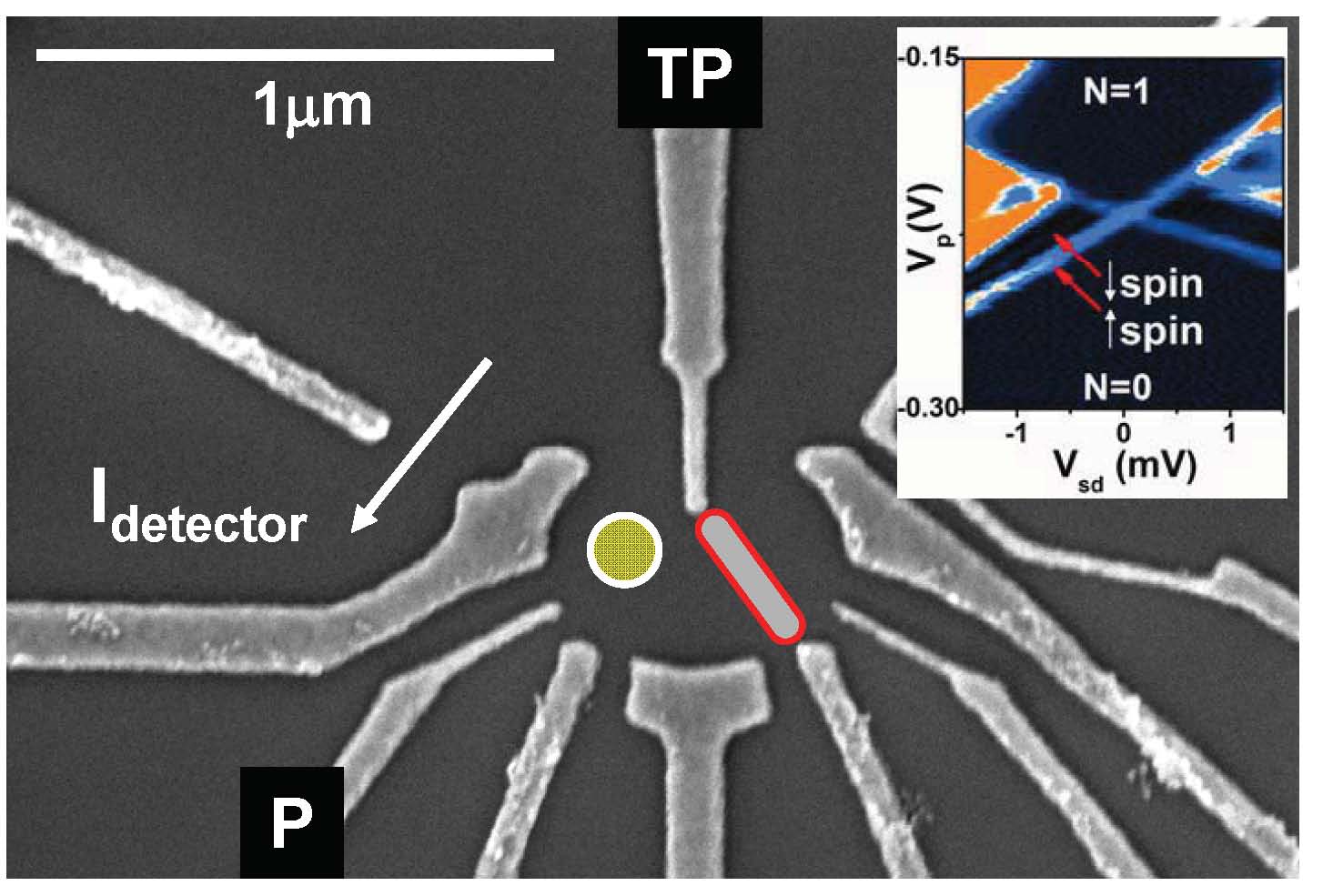}
\caption{SEM image of a device with a similar gate layout. The quantum dot is schematically shown (circle) as well as the barrier which is completly pinched-off (red bar). Electrons can therefore tunnel only between the dot and the left lead. A capacitively coupled quantum point contact on the left is used as a charge detector by  monitoring the current $I_{detector}$. The plunger gate used to control the chemical potential of the dot as well as the top plunger gate controlling the tunnel barrier between the dot and the lead are labeled P and TP respectively. Inset: Transport spectroscopy of the quantum dot in a 10 T magnetic field applied parallel to the plane of the 2DEG. The spin up and spin down states are observed as the ground and first excited states, respectively, in the N=0 to N=1 transition, where N is the number of electrons in the quantum dot.}
\label{fig:1}
\end{center}
\end{figure}

\section{Experiment}
\label{}

Figure 1 shows an sacanning electron micrograph (SEM) image of a device with a similar layout to the one used in this experiment. The device was fabricated on a high mobility AlGaAs/GaAs heterostructure. To reduce switching noise issues, both bias cooling\cite{Pioro-Ladriere2005} and a global gate\cite{Buizert2008} were employed. The results in this paper were obtained by tuning the gates to  isolate a single dot within the device, schematically shown in figure 1. The quantum dot was tuned down to a single electron. For the spin readout sequence, a parallel magnetic field of 10 T was employed to separate the spin qubit's up and down states. These states were observed in transport spectroscopy measurements as the ground and first excited states (inset figure 1). The tunnel barrier between the quantum dot and the right side was set to prevent any tunnelling, while the tunnel barrier between the quantum dot and the left lead was calibrated and tuned during the course of these experiments. The experiments were performed using time resolved charge detection techniques using a quantum point contact (QPC) on the left side of the device as shown. Measurements were made on a dilution refrigerator using a standard high frequency set up.

\section{Results and discussion}
\label{}

The tunnel barrier is calibrated using two techniques. In the first technique, the chemical potential of the lead and the dot are matched and the statistics of occupation fluctuations are used to determine average tunnelling times. Figure 2(a) shows three traces where the chemical potential of the dot is set above, on resonance (i. e. on the Coulomb blockade peak) and below the Fermi energy of the lead by tuning it with gate P. When an electron enters (leaves) the dot, a decrease (increase) in current is measured in the charge detector. If the dot is mainly empty (full), i. e. the chemical potential is above (below) the Fermi energy, the signal measured in the detector is mainly on the high (low) level. When the dot is in resonance with the lead, the signal measured is, on average, in the high and low levels almost equally. This technique allows us to set the dot in resonance with the lead under different potential conditions, for example, when tuning the tunnel barrier to the lead with the top plunger gate. Figure 2(b) shows different time traces where the tunnelling barrier was tuned with gate TP. For a high tunnel barrier, a small number of events is observed (top trace). As the tunnel barrier is decreased by applying a less negative voltage on the top plunger gate, the number of events within the same time scale increases. To illustrate the full counting statistics technique, figure 2(c) shows an example of time resolved detection for a fixed tunnel barrier. In order to extract the tunnelling times, the trace is analysed by establishing a threshold value of the current measured above which the quantum dot is empty and below which the dot contains one electron. A binary trace (top trace in the figure) is then generated, and by analysing the time an electron spends in and out of the quantum dot the tunnelling times   in and  out are extracted. For the top plunger gate setting used in the data shown in figure 2(c), the tunnelling times obtained were   $\tau_{in}$ = 790 $\mu$s and $\tau_{out}$ = 650 $\mu$s, which are very similar because the quantum dot is in resonance with the lead.

\begin{figure}[htb]
\begin{center}
\includegraphics*[scale=0.9]{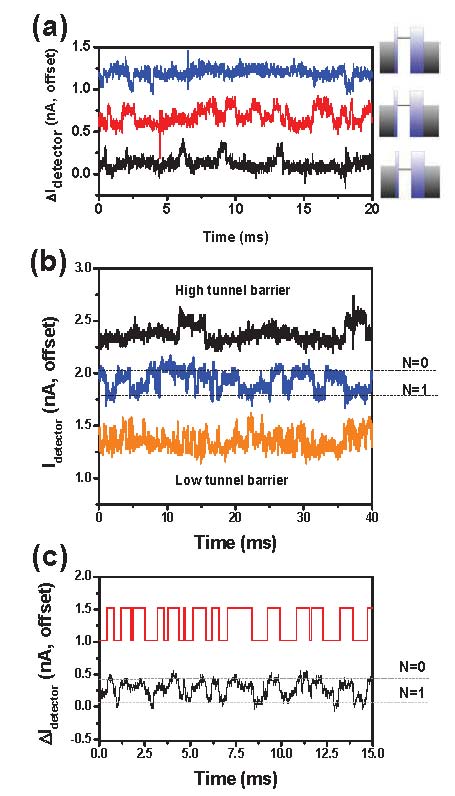}
\caption{(a) Time resolved charge detection signal for different chemical potential settings of the quantum dot. The top, center and bottom traces correspond, respectively, to the chemical potential lying above, on resonance and below the Fermi energy of the lead, sketched on the right side. Single events are picked up by the charge detector as electrons enter and leave the dot. (b) Time resolved charge detection for different tunnel barriers controlled by the top plunger gate. (c) Extraction of the tunnelling time by full counting statistical analysis of single events measured. The top curve is a binary representation of a time trace (bottom trace) which is used to do the tunnelling time analysis. For the particular top plunger gate voltage used we obtain tunnelling times $\tau_{in}$ = 790 $\mu$s and $\tau_{out}$ = 650 $\mu$s.}
\label{fig:2}
\end{center}
\end{figure}

A second approach to extract tunnelling times, which is more suitable for multiple dot devices, is shown in figure 3. The dot is initialised in the empty configuration. A two pulse sequence to the plunger gate is then employed to stochastically fill and then empty the quantum dot.  This is schematically pictured in figure 3(a). The indirect effect of the pulses can be seen as a rise (as the plunger gate voltage is made less negative) and fall (as the plunger gate voltage is made more negative) in the charge detector signal. In addition, a smaller drop in signal is observed when the electron enters the dot and an equivalent rise in signal is seen when it exits. Since this process is stochastic, the exact moment of electron entry and exit is different in every measurement, two of which are shown in figure 3(b). To extract  $\tau_{in}$ and $\tau_{out}$, 500 traces are averaged, leading to an exponential decay during the filling and emptying stages of the measurement. The result is shown in figure 3(c). Using this method, tunnelling times ($\tau_{in}$ = 720 $\mu$s and $\tau_{out}$ = 650 $\mu$s for this tunnel barrier voltages) were obtained for a particular top gate voltage setting. The difference in the two times reflects the shape of the barrier potential and the two different voltage settings in the pulse sequence for tunnelling in and out. The measurements are repeated for different top gate voltage settings. Figure 4 plots the tunnelling times vs. the top gate voltage value confirming our control over the tunnel time for around three orders of magnitude.

\begin{figure}[htb]
\begin{center}
\includegraphics*[scale=0.45]{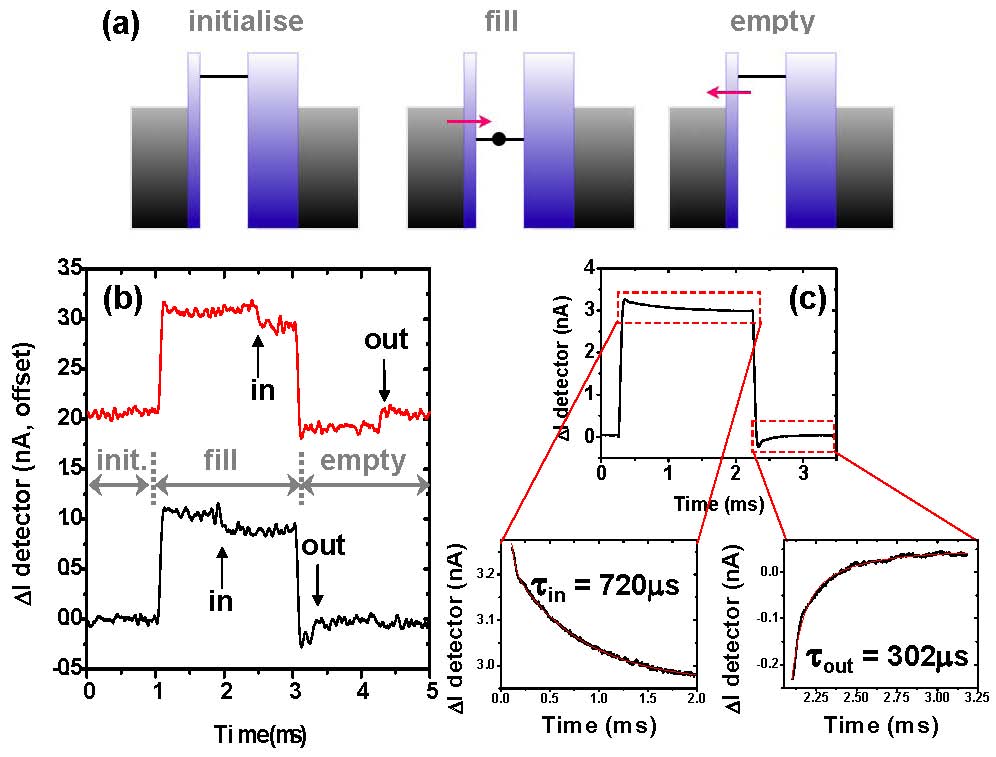}
\caption{(a) Single electron injection-ejection sequence. The quantum dot is first emptied (initialisation). Then the plunger gate is pulsed so that the chemical potential of the dot is below the fermi energy of the lead and an electron is allowed to fill the dot. Finally, after the filling time the gate is pulsed once again to empty the dot. (b) Two example traces showing single-shot measurements of single in-out events measured by the charge detector, showing the stochastic aspect of the tunnelling events. (c) Averaged curve of 500 individual measurements. Insets: The analysis of the exponential decay during the filling and emptying times leads to the extraction of the tunnelling times $\tau_{in}$ = 720 $\mu$s and $\tau_{out}$ = 302 $\mu$s.}
\label{fig:3}
\end{center}
\end{figure}

\begin{figure}[htb]
\begin{center}
\includegraphics*[scale=0.5]{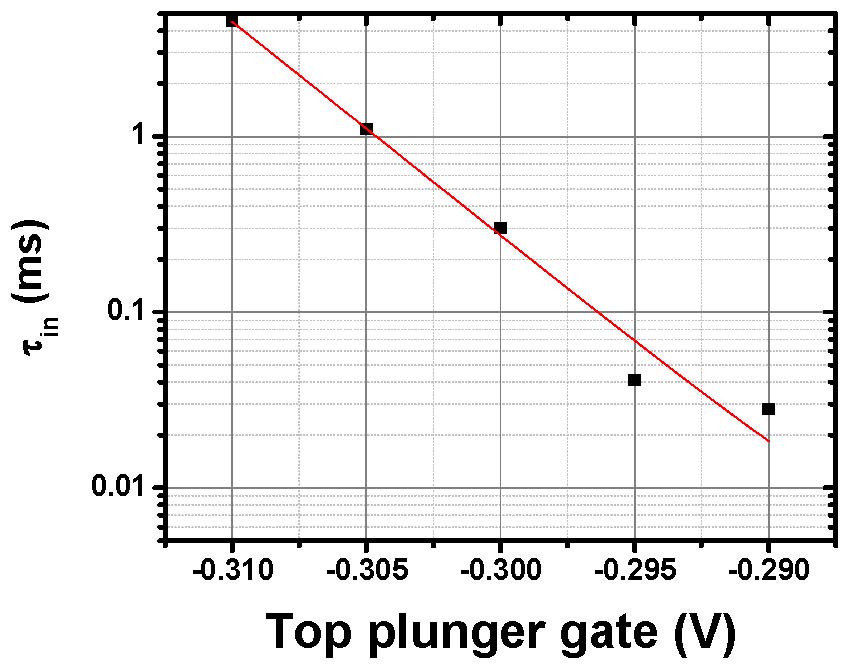}
\caption{Calibration of the tunnelling time $\tau_{in}$ as a function of the top plunger gate. The straight line is a fit using an exponential dependence on the top plunger gate voltage.}
\label{fig:4}
\end{center}
\end{figure}

\begin{figure}[h]
\begin{center}
\includegraphics*[scale=0.75]{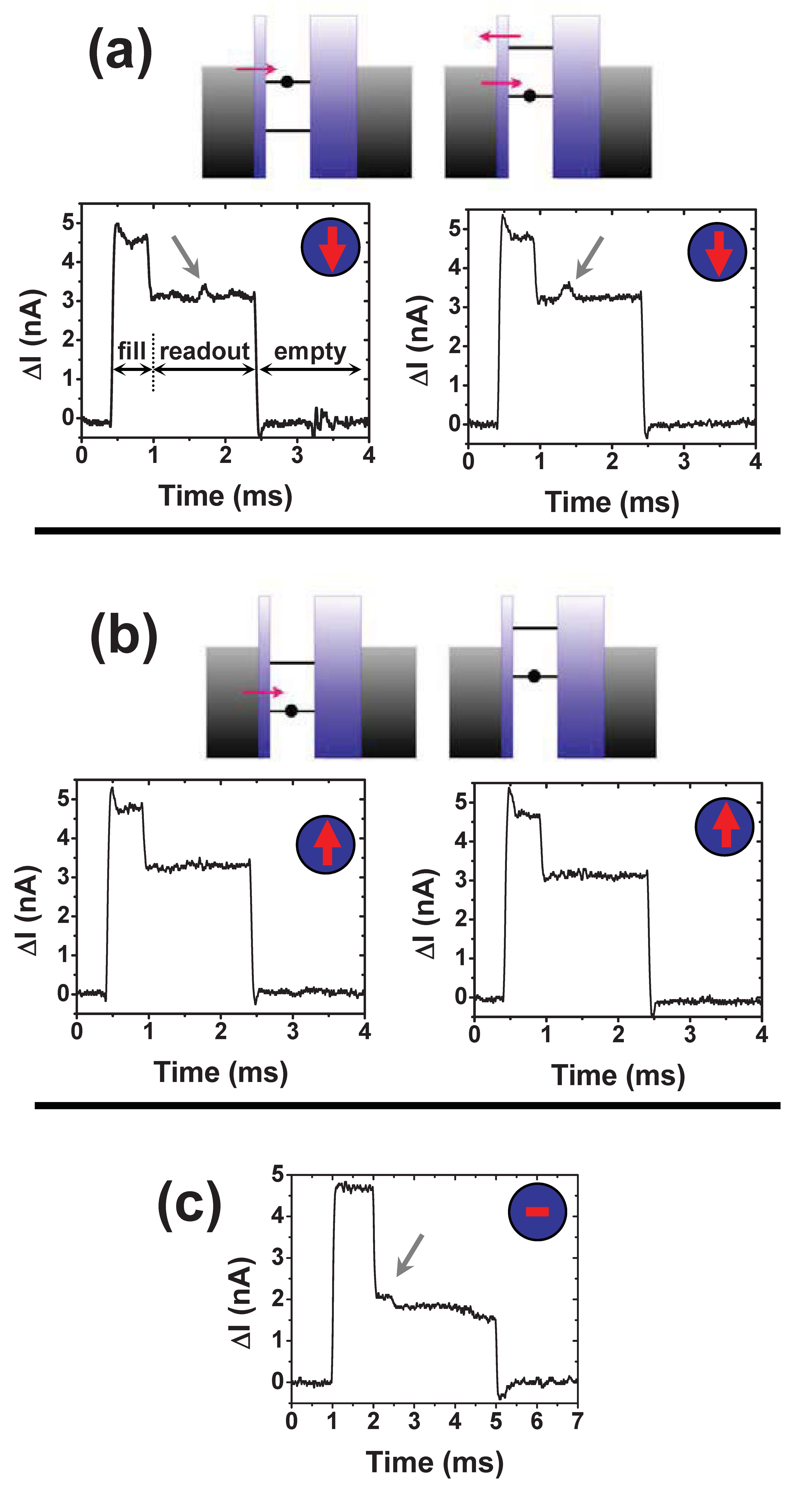}
\caption{Single-shot single electron spin readout experiment based on the Delft protocol. (a) Schematics of the fill and read-out sequences of the protocol in the event of a spin down detection. The charge detector measures an increase and a decrease of current during the readout time as an electron with spin down is ejected and then replaced by a spin up electron, showed by the arrows in the two example traces. (b) Spin up detection. No events are measured in the readout time because the spin up electron is not ejected, as its energy level lies below the Fermi energy of the lead. (c) Example of an invalid measurement, the tunnelling time of the electron exceeds the filling time of the protocol.}
\label{fig:5}
\end{center}
\end{figure}

Using the averaging technique, we tuned the tunnel barrier to perform the single shot spin readout experiment. We had previously shown that spin polarized edge states could be used to readout a single spin utilizing spin blockade effects\cite{Ciorga2000}. Recently, however, a more elegant protocol has been introduced by Elzerman \textit{et al.}\cite{Elzerman2004}. To run this Delft protocol on our device, we first applied a parallel field of 10 T to define our spin up an down qubit states (see inset figure 1). We then introduced the required additional "readout" step in our pulse sequence. The fill step now adds a single electron of unknown spin to the quantum dot. The readout step positions the Fermi energy of the lead between the two spin states. The tunnel barrier to the lead is set to be shorter than the spin relaxation time so that the electron is allowed to enter the dot quickly before it relaxes to the ground state. While ultimately the spin up ground state will be occupied, the charge detector response will be different depending on whether the initial spin state is up or down. If the initial occupied state is spin down then the spin up state becomes occupied by a two step process. First the electron exits and then a spin up re-enters the dot. On the other hand, if the initial occupied state was spin up, then it just remains in the dot since the spin up state has a lower energy than Fermi energy of the lead and hence the electron cannot escape. During the readout step, the charge detector picks up whether an electron exited and another re-entered the dot. If this is observed, then, following the original protocol\cite{Elzerman2004}, we can declare that the original electron was a spin down one. If the readout step does not detect such a process, then we can declare the original electron as a spin up electron. Figures 5 (a) and (b) show schematically the fill and readout processes for each spin as well as example measurements. Clearly in such a stochastic process there are many opportunities for error. One such example is given in figure 5(c). In this case no electron entered during the 'fill' step but an electron entered during the readout stage. From the level configuration during the readout step, we can identify this as a spin up electron.

\begin{figure}[htb]
\begin{center}
\includegraphics*[scale=0.5]{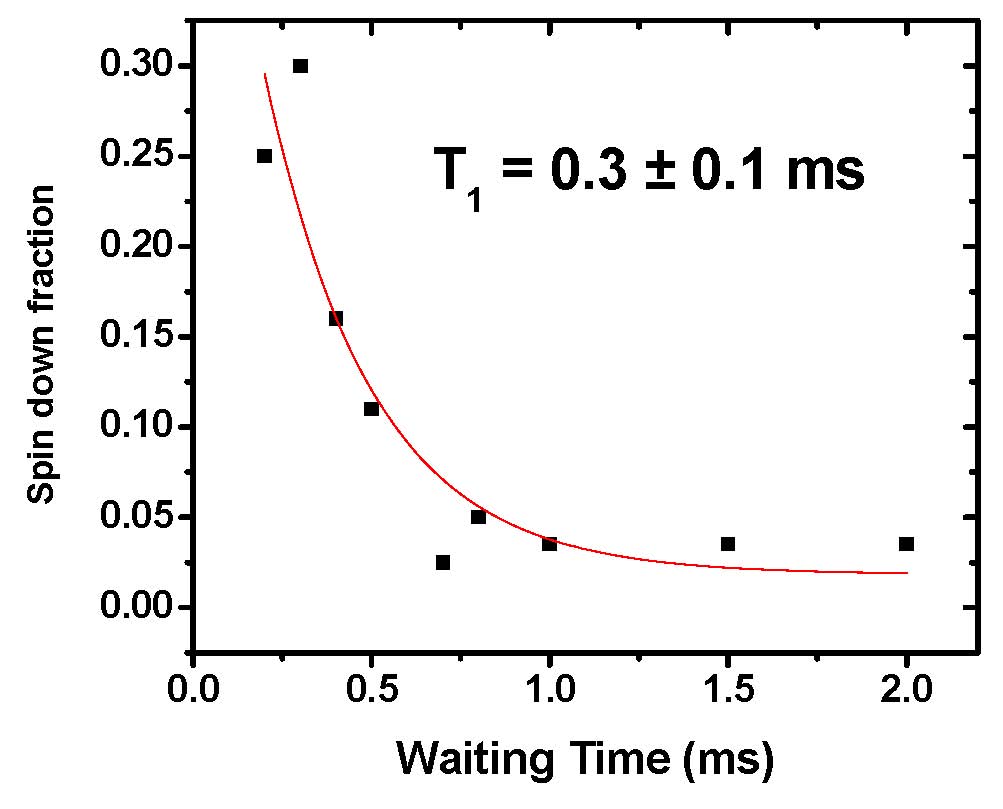}
\caption{Extraction of the spin relaxation time $T_1$ by statistical analysis obtained from the spin down fraction of the measured events as a function of the waiting time before the readout. The obtained relaxation time is $T_1 = 300 \mu s$.}
\label{fig:6}
\end{center}
\end{figure}

To obtain the spin relaxation time T$_1$, we measure the spin down fraction as a function of waiting time prior to readout. This fraction decreases with longer waiting time due to the intrinsic relaxation of the spin down electrons to the spin up ground state. The result of this analysis is shown in figure 6, where an exponential dependence on the waiting time is obtained. This analysis leads to a spin relaxation time T$_1$ = 300 $\mu$s.

\section{Summary}
\label{}

We have demonstrated the control of tunnelling times in and out with both full counting statistics and averaging methods. We have used these techniques to adjust the tunnel barrier between a single quantum dot and the lead in order to run the Delft single shot spin readout protocol in a single dot device. We have obtained a T$_1$ time of 300 $\mu$s at 10 T.

\section{Acknowledgements}
\label{}

We would like to thank S. Branchaud for useful discussions as well as Z.Wasilewski and J.Gupta for the high mobility wafer. A.S.S, and J.K. acknowledge assistance from NSERC and Quantumworks. A.S.S. also acknowledges assistance from CIFAR.

\end{document}